\documentclass[aip,amsmath,amssymb,rsi,groupedaddress]{revtex4-1}
\usepackage{graphicx}
\usepackage{epsfig}
\usepackage{amsmath}
\usepackage{graphics}
\usepackage{epsfig}
\usepackage{yfonts}
\usepackage{amsmath}
\usepackage{amsfonts}
\usepackage{amssymb}
\usepackage{xcolor,cancel}

\begin{document}

\title{Nonequilibrium Green's function theory for nonadiabatic effects in quantum electron transport }
\author{Vincent F. Kershaw} 
\author{Daniel S. Kosov}
\affiliation{College of Science and Engineering, James Cook University, Townsville, QLD, 4811, Australia 
}


\begin{abstract}
We develop nonequilibribrium Green's function based transport theory, which includes effects of  nonadiabatic nuclear motion  in the calculation of the electric current in molecular junctions. Our approach is based on the separation of slow and fast timescales in the equations of motion for the Green's functions by means of the Wigner representation. Time derivatives with respect to central time serves as a small parameter in the perturbative expansion
enabling the computation of nonadiabatic corrections to molecular Green's functions. 
Consequently, we produce series of analytic expressions for non-adiabatic electronic Green's functions (up to the second order in the central time derivatives); which depend not solely on instantaneous molecular geometry but likewise on nuclear velocities and accelerations. Extended formula for electric current is derived which accounts for the non-adiabatic corrections. This theory is concisely illustrated by the calculations on a model molecular junction.
\end{abstract}

\maketitle

\section{Introduction}
The basic building block for molecular electronics is a single-molecule junction: a molecule chemically linked to two macroscopic metal electrodes.\cite{moletronics} To date the major achievements in molecular electronics have been the development of a fundamental understanding of quantum transport in single molecules and ways to control it.  Its major shortcomings have been associated with the inability to physically control nuclear geometry in molecular devices and related to it, the variability of molecular junction electric characteristics. The absence of the structural rigidity, and the resulting strong interplay between electronic and nuclear dynamics are  the most distinctive features of molecular junctions in comparison to silicon based devices. 

Molecular-sized systems will always experience nuclear motion caused by  current, which in turn will be influenced by dynamical conformational changes. Coupling between nuclear and electronic degrees of freedom has lead to a variety of new transport phenomena such as negative differential resistance, Franck-Condon blockade, current induced chemical reactions, and other interesting effects.\cite{galperin09,Ioffe:2008aa,PhysRevB.83.115414,galperin05,zazunov06,reimers06,dzhioev11,catalysis12,fcblockade05,hartle10,thoss11,thoss14}   A multitude of theories of various degrees of sophistication have been proposed to treat vibrationally coupled electron transport: master equations,\cite{PhysRevB.69.245302,fcblockade05,PhysRevB.83.115414,may02,PhysRevB.94.201407,segal15,kosov17-wtd,kosov17-nonren}
path integral,\cite{rabani08} Fock-Liouville space superoperators,\cite{dzhioev14,dzhioev15} scattering theory, \cite{ness05,PhysRevB.70.125406,peskin07,doi:10.1063/1.3231604}  nonequilibrium Green's functions,\cite{caroli72,ryndyk06,dahnovsky:014104,galperin06,ryndyk07,hartle08,rabani14,hartle15,PhysRevB.75.205413} and multilayer multiconfiguration time-dependent Hartree theory.\cite{wang11,wang13,wang13b,Wang2016117}
Yet despite many successful applications, these theories suffer from the same limitation;  the amplitudes of current induces nuclear motion must be small (vibrations around equilibrium molecular geometry). 

 In this paper, we develop  nonequilibrium Green's function based transport theory while taking into account the non-adiabatic effects of nuclear motion.  
While the approach taken remains perturbative, it is not based on the assumption that the amplitude of nuclear motion is either small or harmonic; nor is it assumed that the coupling between nuclear and electronic degrees of freedom is small.  The small parameter in our theory is the velocity of nuclear motion - we assume that the characteristic timescale for conformational changes is larger in comparison to the electronic tunnelling time. This leads to the development of an extended formula for electric current in terms of adiabatic molecular Green's functions, the nuclear gradients of molecular orbital energies, nuclear velocities, and nuclear accelerations.

The outline of the paper is as follows. Section II contains the main theoretical results:  equations of motion for Green's functions in Wigner representation, perturbative solutions of these equations via separation of fast and slow timescales, and the derivation of nonadiabatic corrections to the expression for  electric current. In section III, we apply the proposed theory to electron transport through a single resonant-level molecular junction, then analyse numerically and analytically the role of nuclear motion on molecular junction conductivity. Section IV gives conclusions and  summary of the main results.
We use atomic units in the derivations throughout the paper ($\hbar= |e|= m_e= 1$).

\section{Theory}
\subsection{Hamiltonian, Green's functions and self-energies}
The aim of this section is to present the Hamiltonian, to give  definitions of the Green's functions and self-energies, and to introduce basic notations  which will be used throughout 
the paper.
 
The single-molecular junction is a  molecule connected to macroscopic  leads. 
The system is described by the following Hamiltonian
\begin{equation}
H=H_M + H_L + H_R + H_{LM} + H_{RM},
\end{equation}
where $H_M$ is the  molecular Hamiltonian, $H_L$ is the Hamiltonian for the left lead, $H_R$ is the Hamiltonian for the right lead,  $H_{LM}$ and $H_{RM}$  are interactions between the molecule and left and right leads, respectively. 
We assume that the molecule-lead interactions,  $H_{LM}$ and $H_{RM}$, do not depend on the molecular geometry. This is not critical approximation from the practical point of view, since the partitioning of the system on the "molecule" and leads is somehow arbitrary - we can always include   physically important part of lead-molecule interface into the definition of the "molecule" (for example, two or three layers of electrode metal atoms can be considered as  part the extended "molecule" in molecular junction).
The molecular Hamiltonian  is explicitly time-dependent through the dependence on nuclear positions
\begin{equation}
H_M= \sum_{ij} h_{ij} (\mathbf R(t)) d^\dag_i d_j,
\end{equation}
where the multidimensional vector $\mathbf R(t)$ describes  the molecular geometry  (positions of nuclei) at time $t$ and $d^\dag_i (d_i)$ creates (annihilates) electron in molecular orbital $i$. Here  $h_{ij}$ are matrix elements of the single-particle molecular Hamiltonian computed in some basis. These matrix elements are readily available from density-functional theory, tight-binding or Hartree-Fock molecular electronic structure calculations.  
We assume from the outset that nuclei are classical particles. We omit the classical part of the Hamiltonian (nuclear kinetic energy and  nuclei-nuclei repulsion), since it does not contribute to the equations for electronic Green's functions. 

The left and right leads of the molecular junction are modelled as macroscopic reservoirs of noninteracting electrons
\begin{equation}
H_L =  \sum_{l} \epsilon_{l} a^\dag_{l} a_{l}
\end{equation}
and
\begin{equation}
H_R =  \sum_{r} \epsilon_{r} a^\dag_{r} a_{r},
\end{equation}
where $a^\dagger_{l/r}$   creates  an electron in the single-particle state $l/r$  with energy $\epsilon_{l/r}$ of the left/right lead  and $a_{l/r}$ is the corresponding electron  annihilation operator.   The left lead-molecule  and right lead-molecule couplings are described by the tunneling interaction
\begin{equation}
H_{LM}=  \sum_{li } ( t_{l  i}   a^\dag_{l} d_i  +  t^*_{li}  d_i^\dag a_{l })
\end{equation}
and
\begin{equation}
H_{RM}=  \sum_{ri } ( t_{r  i}   a^\dag_{r} d_i  +  t^*_{ri}  d_i^\dag a_{r }),
\end{equation}
where $t_{li}$  and $t_{ri}$ are the tunnelling amplitudes between  leads and molecule single-particle states.

We define exact (nonadiabatic, computed with full time-dependent Hamiltonian  along given nuclear trajectory $\mathbf R(t)$) retarded, advanced, and lesser Green's functions as
\begin{equation}
{\cal G}_{ij}^R(t,t') = -i \theta(t-t') \langle \{d_i (t), d^\dag_j (t')\} \rangle,
\end{equation}
\begin{equation}
{\cal G}_{ij}^A(t,t') = \Big({\cal G}_{ji}^R(t',t) \Big)^*
\end{equation}
and
\begin{equation}
{\cal G}_{ij}^<(t,t') = i \langle d^\dag_j (t')d_i (t) \rangle.
\end{equation}

The self-energies of leads are not affected by the time dependent molecular Hamiltonian and they are defined in standard way.\cite{haug-jauho}  Left and right retarded self-energies are
\begin{equation}
{\Sigma}_{L ij}^R(t,t') = -i \theta(t-t') \sum_{l} t^*_{l i}  e^{-i \epsilon_{l} (t-t')} t_{l  j}
\end{equation}
and
\begin{equation}
{\Sigma}_{R ij}^R(t,t') = -i \theta(t-t') \sum_{r} t^*_{r i}  e^{-i \epsilon_{r} (t-t')} t_{r j}.
\label{sigmaRt}
\end{equation}
The advanced  and retarded self-energies are related to each other  via Hermitian  conjugation:
\begin{equation}
{\Sigma}_{L/R ij }^A(t,t') = \Big( {\Sigma}_{L/R ji}^R(t',t) \Big)^*.
\label{sigmaAt}
\end{equation}
The lesser self-energies are
\begin{equation}
{\Sigma}_{L ij}^<(t,t') =2 \pi i \sum_{l} t^*_{l i}  f_L(\epsilon_l) e^{-i \epsilon_{l} (t-t')} t_{l  j}
\end{equation} 
and 
\begin{equation}
{\Sigma}^<_{R ij} (t,t') = 2 \pi i  \sum_{r} t^*_{r i}  f_R(\epsilon_r) e^{-i \epsilon_{r} (t-t')} t_{r j},
\label{sigmaRt}
\end{equation} 
where $f_L$ and $f_R$ are the Fermi-Dirac occupation numbers for the left and right leads.
The total self-energies are the sum of contributions from the left and right leads
\begin{equation}
{\Sigma}^{R,A,<}_{ij}(t,t') =  {\Sigma}^{R,A,<}_{L ij}(t,t')+ {\Sigma}_{R ij}^{R,A,<}(t,t').
\end{equation}

The retarded self-energies in the energy domain  are defined in the usual way as a Fourier transformation of time domain self-energies defined above:
\begin{equation}
\Sigma_{L/R ij}^R(\omega)= \Delta_{L/R ij }(\omega)  -\frac{i}{2} \Gamma_{L/R ij}(\omega),
\end{equation}
where the level-width functions are
\begin{equation}
 \Gamma_{L ij } (\omega) = 2 \pi \sum_{l} \delta(\omega-\epsilon_l)  t^*_{l i} t_{l j}
\end{equation}
 and
\begin{equation}
  \Gamma_{R ij }(\omega) = 2 \pi \sum_{r} \delta(\omega-\epsilon_r)  t^*_{ri} t_{r j}.
\end{equation}
The level-shift functions $\Delta_{L/R ij }(\omega)$ can be computed from $\Gamma_{L/R ij}(\omega)$  via Kramers-Kronig relation.\cite{haug-jauho} 
The advanced and lesser  self-energies are computed from the retarded self-energy as
\begin{equation}
\Sigma_{L/R ij}^A(\omega) = (\Sigma_{L/R ji}^R(\omega))^*
\end{equation}
 and 
\begin{equation}
 \Sigma_{L/R ij}^<(\omega)= f_{L/R}(\omega)\left( \Sigma_{L/R ij}^A(\omega) - \Sigma_{L/R ij}^R(\omega)\right)= i f_{L/R}(\omega) \Gamma_{L/R ij}(\omega).
\end{equation}

\subsection{Separation of timescales in equations of motion and nonadiabatic corrections to the retarded Green's function}

We begin with the equation of motion for the retarded Green's function\cite{haug-jauho}
\begin{equation}
[ i\partial_t  - h(t) ] {\cal G }^{R}(t,t') = 
 \delta(t-t') +  \int^{+\infty}_{-\infty} dt_1 \Sigma^{R}(t,t_{1}){\cal G }^{R}(t_{1},t') .
 \label{eom1}
\end{equation}
Note that the Green's functions, self-energies and Hamiltonian $h$ are matrices in the molecular orbital space but we will not write explicitly molecular orbital indexes in the equations for simplicity of presentation.  The Green's function ${\cal G }^{R}$, which satisfies this equation, is  nonadiabatic and  exact.

We define central time $T$ and relative  time $\tau$ for  the Green's function ${\cal G }^{R}(t,t')$ as
\begin{equation}
T = \frac{1}{2}(t+t'), \;\;\;
\tau = t - t',
\end{equation}
and introduce the Wigner representation of the Green's function
\begin{equation}
{\cal \widetilde G}^R(\omega,T) = \int^{+\infty}_{-\infty}  d\tau \; e^{i \omega \tau} {\cal G}^R (t,t') .
\end{equation}
The inverse transformation from the Wigner representation to time domain is
\begin{equation}
{\cal G}^R (t,t')= \frac{1}{2 \pi} \int^{+\infty}_{-\infty}  d\omega \; e^{-i \omega \tau}  {\cal \widetilde G}^R(\omega,T).
\end{equation}
The equation of motion (\ref{eom1}) in its Wigner representation is given by 
\begin{equation}
\left[ \omega  +\frac{i}{2}\partial_T  - e^{\frac{1}{2i}\partial_{\omega}^{{\cal G }} \partial_{T}^{h}}h (T) \right] \widetilde{{\cal G }}^{R}(T,\omega) \\
= I +  e^{-\frac{1}{2i}\partial_{\omega}^{\Sigma} \partial_{T}^{{\cal G }}} {\Sigma}^{R}(\omega)\widetilde{{\cal G }}^{R}(T,\omega).
\label{eomW}
\end{equation}
Here $\partial^\Sigma$, $\partial^{\cal G }$ and $\partial^{h }$ mean the derivatives acting on self-energy $\Sigma^R$, Green's function $\widetilde{{\cal G }}^{R}$ and Hamiltonian matrix $h$, respectively. $I$ is the identity matrix.
The details of the derivation of (\ref{eomW}) are given in Appendix A.

So far we have not made any approximations. The equation of motion (\ref{eomW}) describes  the exact   nonadiabatic evolution of the retarded Green's function; it must be solved along a given trajectory $\mathbf R(t)$ of nuclear coordinates. The advantage of the use of  Wigner representation is that fast and slow timescales are easily identifiable. For the Green's function ${\cal G}^{R}(t,t')$, the slow nuclear motion implies that ${\cal G}^{R}(T+\tau/2,T-\tau/2)$ varies slowly with the central time $T$, but oscillates fast with the relative time $\tau$.
 This is because the central time dependence in the Green's function can only originate from the explicit time-dependence of the electronic Hamiltonian 
(that is the dependence of $h(\mathbf R(t))$ on a nuclear trajectory) whereas dependence on the relative time is always present for both static and time dependent Hamiltonians.

We assume that  the variations of the retarded Green's function with respect to central time can be considered small and as such the exponential operators $e^{\frac{1}{2i}\partial_{\omega}^{\widetilde{G}} \partial_{T}^{R}}$ and $e^{-\frac{1}{2i}\partial_{\omega}^{\Sigma}\partial_{T}^{{\cal G }}} $ are expanded as a convergent finite Maclaurin series. Estimated at resonance,  the  first and second derivative terms of the retarded Green's function are of order $ \dot{x} \partial_{T / \omega} \widetilde{{\cal G}}^{R} \sim (\frac{\Omega}{\Gamma})$ and $ \ddot{x} \partial_{T / \omega}^{2} \widetilde{{\cal G}}^{R} \sim  \dot{x}^2 \partial_{T / \omega}^{2} \widetilde{{\cal G}}^{R}\sim (\frac{\Omega}{\Gamma})^{2}$ respectively, where $\Omega$ is the characteristic frequency of the molecular vibrations. If we assume that electronic tunneling time is faster than the period of the molecular vibrations (that means $\Gamma \gg \Omega$), we can restrict the Maclaurin series for   $e^{\frac{1}{2i}\partial_{\omega}^{\widetilde{G}} \partial_{T}^{R}}$  and $e^{-\frac{1}{2i}\partial_{\omega}^{\Sigma}\partial_{T}^{{\cal G }}} $ to the first few terms.

To expand the exponential operators $e^{\frac{1}{2i}\partial_{\omega}^{\widetilde{G}} \partial_{T}^{R}}$  and $e^{-\frac{1}{2i}\partial_{\omega}^{\Sigma}\partial_{T}^{{\cal G }}} $ in equation of motion (\ref{eomW})  we first form a $3N$ dimensional vector, which describes the geometry of the molecule in the junction
\begin{equation}
\mathbf R = (\underbrace{ x_1, x_2, x_3}_{\mathbf R_1}, \underbrace{ x_4, x_5, x_6}_{\mathbf R_2}, ....., \underbrace{ x_{3N-2}, x_{3N-1}, x_{3N}}_{\mathbf R_N}).
\end{equation}
The first and second order time derivatives of the Hamiltonian matrix are
\begin{equation}
\partial_T h(\mathbf R) = \sum^{3N}_{\alpha=1}  \dot{ x}_\alpha \frac{\partial h(\mathbf R)}{\partial x_\alpha},
\end{equation}
and
\begin{equation}
\partial^2_T h(\mathbf R) =  \sum_{\alpha=1}^{3N}  \ddot{x}_\alpha \frac{\partial h(\mathbf R)}{\partial x_\alpha} + \sum^{3N}_{\alpha,\beta=1}  \dot{x}_\alpha \dot{x}_\beta  \frac{\partial^2 h(\mathbf R)}{\partial x_\alpha \partial x_\beta}. 
\end{equation}
Introducing  matrices in the molecular orbital space $\Lambda_{\alpha}$ and $\Phi_{\alpha \beta}$ such that 
\begin{equation}
\Lambda_{\alpha}(\mathbf R)=\frac{\partial h(\mathbf R)}{\partial x_\alpha},
\end{equation}
and
\begin{equation}
\Phi_{\alpha \beta}(\mathbf R) =\frac{\partial^2 h(\mathbf R)}{\partial x_\alpha \partial x_\beta},
\end{equation}
we get
\begin{equation}
\partial_T h =  \sum^{3N}_{\alpha=1}  \dot{ x}_\alpha \Lambda_\alpha(\mathbf R),
\end{equation}
and
\begin{equation}
\partial^2_T h= \sum_{\alpha=1}^{3N}  \ddot{x}_\alpha  \Lambda_\alpha(\mathbf R)+ \sum^{3N}_{\alpha,\beta=1} \dot{x}_\alpha \dot{x}_{\beta}  \Phi_{\alpha \beta}(\mathbf R).
\end{equation}
Finally, we use  Einstein convention for summation over repeated Greek indices to express compactly these derivatives as
\begin{equation}
\partial_T h(\mathbf R) =   \dot{ x}_\alpha \Lambda_\alpha (\mathbf R) ,
\end{equation}
and
\begin{equation}
\partial^2_T h(\mathbf R)
=  \ddot{x}_\alpha  \Lambda_\alpha (\mathbf R)+  \dot{x}_\alpha \dot{x}_{\beta}  \Phi_{\alpha \beta} (\mathbf R).
\end{equation}
We would like to emphasise again that $h$ and, therefore $\Lambda$ and $\Phi$, are themselves matrices which depend on two molecular orbital indices.

We are ready now to introduce the approximations into equation of motion (\ref{eomW}) and then  solve it.
The zeroth order expansion of the exponents in (\ref{eomW}) results into the standard adiabatic equations of motion, the first order expansion gives corrections linear in 
nuclear velocities \cite{Bode12,catalysis12,galperin15,subotnik17}  (it vanishes when it is averaged over nuclear motion), only the second order terms in the exponent expansions contain  nonvanishing contributions.
Keeping the terms  up to the second order in (\ref{eomW})  we obtain the following  equations of motion:
\begin{eqnarray}
\left[ \omega  +\frac{i}{2}\partial_T  - h-  \frac{1}{2i}\dot{ x}_\alpha \Lambda_\alpha \partial_{\omega} + \frac{1}{8}  \ddot{x}_\alpha  \Lambda_\alpha \partial^{2}_{\omega} \right.  \left. + \frac{1}{8}  \dot{x}_\alpha \dot{x}_{\beta}  \Phi_{\alpha \beta}  \partial^{2}_{\omega} \right]
\widetilde{{\cal G }}^{R}
\nonumber
\\
=  I +  \Sigma^{R}\widetilde{{\cal G }}^{R} -\frac{1}{2i} \partial_\omega \Sigma^R \partial_T \widetilde{{\cal G }}^{R}  - \frac{1}{8} \partial^2_\omega \Sigma^R \partial^2_T \widetilde{{\cal G }}^{R}.
\label{eom-final}
\end{eqnarray}
Note that we now omit the Green's and self-energy function arguments $(T,\omega)$ in the interest of brevity.
This equation  can be resolved analytically by a series expansion of the  Green's function  in the power of the small parameter up to the second order
\begin{equation}
\label{expansion}
\widetilde {\cal G }^R = \widetilde {\cal G }^R_{(0)}+ \widetilde{\cal G}^R_{(1)} + \widetilde{\cal G}^R_{(2)} + ....
\end{equation}
Substituting (\ref{expansion}) into  (\ref{eom-final}) we obtain  system of three equations based on order of the small parameter as 
\begin{eqnarray}
\label{eqforg0}
\left[ \omega  - h - {\Sigma}^{R} \right] \widetilde{\cal G}_{(0)}^{R} &=& I ,
\\
\left[ \omega  - h - {\Sigma}^{R} \right] \widetilde{\cal G}_{(1)}^{R} &=&   \frac{1}{2i} \left[ \partial_{T} + \dot{ x}_\alpha \Lambda_\alpha \partial_{\omega} \right] \widetilde{\cal G}_{(0)}^{R} 
-\frac{1}{2i} \partial_\omega {\Sigma}^{R} \partial_T \widetilde{\cal G}_{(0)}^{R},
\label{eqforg1}
\\
\nonumber
\left[ \omega  - h - {\Sigma}^{R} \right] \widetilde{\cal G}_{(2)}^{R} &=& - \frac{1}{8}  \ddot{x}_\alpha  \Lambda_\alpha \partial^{2}_{\omega} \widetilde{\cal G}_{(0)}^{R} 
- \frac{1}{8}  \dot{x}_\alpha \dot{x}_{\beta}  \Phi_{\alpha \beta}  \partial^{2}_{\omega} \widetilde{\cal G}_{(0)}^{R} 
+ \frac{1}{2i} \left[ \partial_{T} 
+ \dot{ x}_\alpha \Lambda_\alpha \partial_{\omega} \right] \widetilde{\cal G}_{(1)}^{R} 
\\
 &-& \frac{1}{2i} \partial_\omega {\Sigma}^{R} \partial_T \widetilde{\cal G}_{(1)}^{R}  -\frac{1}{8} \partial^2_\omega {\Sigma}^{R} \partial^2_T \widetilde{\cal G}_{(0)}^{R}.
\label{eqforg2}
\end{eqnarray}
This system of equations is solved one equation  by one starting from the lowest order. We solve the first equation for $\widetilde{\cal G}_{(0)}^{R}$ and substitute the solution into the second equation, then, we solve the second equation  for $\widetilde{\cal G}_{(1)}^{R}$ and substitute the solution into the third equation, and finally solve the third equation for $\widetilde{\cal G}_{(2)}^{R}$.
The solution of  (\ref{eqforg0}) is  
\begin{equation}
\label{g0}
\widetilde{\cal G}^{R}_{(0)} = \left[ \omega  - h- {\Sigma}^{R} \right]^{-1},
\end{equation}
therefore, the zeroth order retarded Green's function in Wigner representation is simply the standard adiabatic retarded Green's function
\begin{equation}
\widetilde{\cal G}^{R}_{(0)} =G^{R}.
\end{equation}
The equation for the first order correction (\ref{eqforg1}) has the following solution
\begin{equation}
\widetilde{\cal G}_{(1)}^{R} = \frac{1}{2i} {G}^{R} \left[ \partial_{T}+ \dot{ x}_\alpha \Lambda_\alpha  \partial_{\omega} \right]  {G}^{R}.
\end{equation}
The derivatives of the adiabatic retarded Green's function can be easily computed
\begin{equation}
\partial_{T} { G}^{R} = \dot{ x}_\alpha {G}^{R} \Lambda_\alpha {G}^{R}, 
\end{equation}
and
\begin{equation}
\partial_{\omega}{G}^{R} = -G^{R} (I-\partial_\omega \Sigma^R)  G^{R}.
\end{equation}
Here we used the following rule to compute derivatives of the inverse matrix: $ \frac{d A^{-1}}{d\lambda} = - A^{-1}  \frac{dA}{d\lambda} A^{-1} $.
Using these derivatives we get
\begin{equation}
\label{g1}
\tilde{\cal G}_{(1)}^{R} =  \frac{1}{2i}\dot{ x}_\alpha G^{R} \left[ (I-\partial_\omega \Sigma^R)  G^{R}, \Lambda_\alpha  G^{R} \right],
\end{equation}
where $\left[.,\; .  \right]$ is the commutator.
The same expression for the first order nonadiabatic correction to the retarded Green's function  was obtained by Bode and others\cite{Bode12} in the study of current induced forces in mesoscopic conductors.

The equation for $\widetilde{\cal G}_{(2)}^{R}$  (\ref{eqforg2}) involves the second order derivatives of the adiabatic retarded Green's function, which are
\begin{equation}
\partial^2_T G^R =  G^R \left\{ \ddot x_\alpha \Lambda_\alpha   + \dot x_\alpha \dot x_\beta  \left( 2 \Lambda_\alpha G^R \Lambda_\beta + \Phi_{\alpha \beta} \right)\right\} G^R
\end{equation}
and
\begin{equation}
\partial^2_\omega G^R =  G^R \left\{ 2 (I-\partial_\omega \Sigma^R)  G^R (I-\partial_\omega \Sigma^R)   + \partial^2_\omega \Sigma^R \right\} G^R.
\end{equation}
The direct substitution of these derivatives  in (\ref{eqforg2}) and rearrangement of the terms  yield
\begin{eqnarray}
\nonumber
 &&\tilde{\cal G}_{(2)}^{R} = - \frac{1}{8} G^R \Big( \ddot{x}_\alpha \Lambda_\alpha + \dot{x}_\alpha \dot{x}_{\beta} \Phi_{\alpha \beta}  \Big) G^R \Big(  2 {\cal A}  G^R  {\cal A}
 + \partial_{\omega}^{2} \Sigma^R \Big) G^R
+ \frac{1}{4} \ddot{ x}_\alpha \Big( G^R \Big)^{2} \Big[ \Lambda_\alpha  G^R, {\cal A} G^R \Big] 
\\
\nonumber
&& + \frac{1}{4} \dot{ x}_\alpha \dot{ x}_\beta G^R \Big[ {\cal A} G^R , \Lambda_\beta  G^R \Big] \Big[ \Lambda_\alpha  G^R, {\cal A} G^R \Big]  
 + \frac{1}{4} \dot{ x}_\alpha \dot{ x}_\beta G^R {\cal A} G^R \Big[ \Big( \Phi_{\alpha \beta} + \Lambda_{\alpha}  G^R \Lambda_\beta \Big)  G^R, {\cal A}  G^R \Big] 
\\
\nonumber
&&+  \frac{1}{4} \dot{ x}_\alpha \dot{ x}_\beta G^R {\cal A}  G^R \Big[ \Lambda_{\alpha} G^R, {\cal A} G^R \Lambda_\beta G^R \Big] 
 - \frac{1}{4} \dot{ x}_\alpha \dot{ x}_\beta G^R \Lambda_{\alpha}  G^R \Big[ \Lambda_{\beta} G^R  {\cal A} G^R, {\cal A}  G^R \Big]  
 \\
 &&- \frac{1}{4} \dot{ x}_\alpha \dot{ x}_\beta G^R \Lambda_{\alpha}  G^R \Big[ \Lambda_{\beta} G^R, \big( \partial_{\omega}^{2}  \Sigma^R +  {\cal A} G^R  {\cal A} \big)  G^R \Big],
 \label{g2}
\end{eqnarray}
where we introduced 
\begin{equation}
{\cal A }=(I-\partial_\omega \Sigma^R).
\end{equation}

\subsection{Electric current}

Having obtained the nonadiabatic corrections to the retarded Green's function, we derive the equation for the electric current. We begin with the general expression  for electric current flowing into the molecule from left/right leads at time $t$:\cite{haug-jauho}
\begin{eqnarray}
J_{L/R}(t) =2 \; \text{Re} \int^{+\infty}_{-\infty}  dt_1
\text{Tr} \left[ {\cal G}^<(t,t_1)\Sigma_{L/R}^A(t_1,t)
+ {\cal G}^R(t,t_1) \Sigma_{L/R}^<(t_1,t) \right],
\end{eqnarray}
where the trace is taken over the molecular orbital indices.
Let us consider the expectation value of the number of electrons in the molecule $N(t) = \langle \sum_i d^\dag_i d _i \rangle$ and apply the continuity equation
\begin{equation}
\frac{d}{dt} N(t) =J_L(t) + J_R(t).
\end{equation}
For arbitrary real number $y$
\begin{equation}
J_L(t) = y J_L(t) + (1-y) [ \frac{d}{dt} N(t) - J_R(t)].
\end{equation}
Assume that couplings to left and right leads are proportional to each other
\begin{equation}
\Gamma_L(\omega) = \lambda \Gamma_R(\omega).
\end{equation}
Choosing $y= 1/(1+\lambda)$ yields
\begin{equation}
J_L(t)   =  \frac{\lambda}{1+\lambda}  \frac{dN}{dt} 
+ \frac{2}{\lambda +1} \text{Re}   \int^{+\infty}_{-\infty} dt_1
 \text{Tr} \left[ {\cal G}^R(t,t_1)  
  \left( \Sigma_L^<(t_1,t) - \lambda \Sigma_R^<(t_1,t) \right) \right] 
  \label{JL1}
\end{equation}
Let us define  the two-time function
\begin{eqnarray}
\nonumber
C(t,t')= \int^{+\infty}_{-\infty}   {\cal G}^R(t,t_1)  \left( \Sigma_L^<(t_1,t') - \lambda \Sigma_R^<(t_1,t') \right)  dt_1,
\label{c}
\end{eqnarray}
which becomes the integral in expression for the current (\ref{JL1}) if we set $t=t'$.
Using (\ref{a9}) and taking into account that the self-energy does not depend on the central time, we obtain
the Wigner representation of (\ref{c}):
\begin{equation}
\widetilde{ C}(T,\omega)=e^{-\frac{1}{2i}\partial_{T}^{G} \partial_{\omega}^{\Sigma}} \widetilde {\cal G}^R(T,\omega)  \left( \Sigma_L^<(\omega) - \lambda \Sigma_R^<(\omega) \right).
\end{equation}
Expanding the exponent and keeping the terms up to the second order, we get
\begin{eqnarray}
\widetilde{ C}(T,\omega)= \left[I -\frac{1}{2i}\partial_{T}^{G} \partial_{\omega}^{\Sigma}   
-\frac{1}{8} (\partial_{T}^{G} \partial_{\omega}^{\Sigma})^2   \right] 
 G^R \left( \Sigma_L^< - \lambda \Sigma_R^< \right)
 \nonumber
 \\
 +\left[I -\frac{1}{2i}\partial_{T}^{G} \partial_{\omega}^{\Sigma}    \right]   \widetilde {\cal G}^R_{(1)}\left( \Sigma_L^< - \lambda \Sigma_R^< \right) +  
 \widetilde {\cal G}^R_{(2)}\left( \Sigma_L^< - \lambda \Sigma_R^< \right).
\end{eqnarray}
Converting  $\widetilde{ C}(T,\omega)$ back to the time domain and letting $t\rightarrow t'$, we obtain the following expression for the current (note that, since $G^R(\omega) \rightarrow 0$ when $\omega \rightarrow \pm \infty$, we perform here the integration by parts to move $\partial_\omega$  and $\partial^2_\omega$ derivatives from the self-energy to the Green's function):
\begin{eqnarray}
J_L(t)   = \frac{\lambda}{1+\lambda}  \frac{dN}{dt} 
+ \frac{1}{\pi(\lambda +1)} \text{Re} \;  \int^{+\infty}_{-\infty} d\omega 
\text{Tr} \Big[ \big( \Sigma_L^< - \lambda \Sigma_R^< \big)
\nonumber
\\
 \times
 ( \Big[I +\frac{1}{2i}\partial_{T} \partial_{\omega}  
-\frac{1}{8} \partial^2_{T} \partial^2_{\omega}   \big] 
 G^R 
 +\big[I +\frac{1}{2i}\partial_{T} \partial_{\omega}   \big]   \widetilde {\cal G}^R_{(1)}
+  
 \widetilde {\cal G}^R_{(2)} \Big) \Big].
\end{eqnarray}
We now average this equation over nuclear velocities. All terms linear in velocities and accelerations disappear, and   $\overline{\langle dN/dt \rangle} =0$ 
(based on physical consideration, if it is not 0, then $\overline{\langle N(t) \rangle}$ cannot be finite). Here the over-line indicates the averaging over nuclear velocities.
\begin{equation}
\overline{ J_L(t)}   =\frac{1}{\pi(\lambda +1)} \text{Re} \;  \int^{+\infty}_{-\infty} d\omega
\text{Tr} \left[ \left( \Sigma_L^< - \lambda \Sigma_R^< \right) \left( 
G^R +  \overline{ \widetilde {\cal G}^R_{(2)} }
-\frac{1}{8} \overline{\partial^2_{T} \partial^2_{\omega}  G^R }
 +\frac{1}{2i} \overline{ \partial_{T} \partial_{\omega}    \widetilde {\cal G}^R_{(1)}  }
\right) \right]  
\end{equation}
Substituting $\lambda$ gives
\begin{equation}
\overline{ J_L(t)}     = \int^{+\infty}_{-\infty}   d\omega
{\cal T}(\omega) (f_L(\omega)-f_R(\omega))  ,
\label{landauer}
\end{equation}
where
\begin{equation}
{\cal T}(\omega) = - \frac{1}{\pi} \text{Tr} \left[\frac{\Gamma_L(\omega) \Gamma_R(\omega)}{\Gamma_L(\omega) + \Gamma_R(\omega)} \; \text{Im} \left( 
G^R +  \overline{ \widetilde {\cal G}^R_{(2)}  }
-\frac{1}{8} \overline{ \partial^2_{T} \partial^2_{\omega}  
 G^R}
 +\frac{1}{2i} \overline{ \partial_{T} \partial_{\omega}    \widetilde {\cal G}^R_{(1)} }
\right) \right].
\label{T}
\end{equation}
Averaged over nuclear velocity distribution quantities  $\overline{ \widetilde {\cal G}^R_{(2)} }$,
$\overline{ \partial^2_{T} \partial^2_{\omega}  G^R }$,  and
$\overline{ \partial_{T} \partial_{\omega}    \widetilde {\cal G}^R_{(1)} }$, which determine ${\cal T}(\omega) $,   are given in Appendix B.
We would like to note here that expression for the current (\ref{landauer},\ref{T}) is general and valid for the case of correlated electrons.
Deceptive in its apparent  similarity with the Landauer formula, ${\cal T}(\omega) $ in  (\ref{landauer}) should not be regarded as the transmission probability for electron with energy $\omega$. The inelastic effects are present in our model (due to the nonadiabatic coupling between electronic and nuclear degrees of freedom) and, therefore, there is no connection between ${\cal T}(\omega) $  (\ref{T}) and the transmission probability.

\section{Example calculations}

To illustrate the proposed theory we consider a model  molecular junction.  The molecule is represented by  a single molecular orbital
with energy $ \epsilon(x(t))$. Molecular orbital energy  $\epsilon(x(t))$ depends on single classical degrees of freedom $x(t)$.
The molecular Hamiltonian is
\begin{equation}
H_M(t) = \epsilon(x(t)) d^\dag d.
\end{equation}
We neglect electron spin here since it is not relevant to our discussion.
The molecule-lead  interaction is treated in the wide-band approximation that means
 the level broadening functions $\Gamma_L$ and $\Gamma_R$ are energy-independent constants (and, therefore, real part of the retarded self-energy $\Delta^R_{L/R}=0$.) Hence,
\[
\partial_\omega \Sigma_L^R=\partial_\omega \Sigma_R^R=0,
\]
and all derivatives of the retarded self-energy disappear from the equations for Green's functions (\ref{g1}, \ref{g2}).
Since the molecular space consists of one orbital,
all Green's functions, self-energies, $\Lambda$, $\Phi$, and $h$ become numbers and consequently all commutators in (\ref{g1}, \ref{g2}) vanish.
Using (\ref{g0},\ref{g1},\ref{g2}) we obtain the following expressions for the retarded molecular Green's function:
Adiabatic (zeroth order) is 
\begin{equation}
G^R= \frac{1}{\omega-\epsilon(x)+i(\Gamma_L + \Gamma_R)/2},
\end{equation}
there is no first order nonadiabatic correction:
\begin{equation}
\tilde{\cal G}_{(1)}^{R} = 0,
\end{equation}
and only  the second order nonadiabatic correction gives the nonvanishing contribution
\begin{equation}
\tilde{\cal G}_{(2)}^{R} = - \frac{1}{4}  \left[\ddot{x}\Lambda (x) + \dot{x}^{2}\Phi (x) \right] \left[ G^{R} \right]^{4}.
\end{equation}

Using these nonadiabatic retarded Green's functions we obtain the following expression for function ${\cal T}(\omega)$ (\ref{T}) that enters the equation for the electric current (\ref{landauer}):
\begin{equation}
{\cal T}(\omega) = - \frac{1}{\pi}  \frac{\Gamma_L \Gamma_R}{\Gamma_L + \Gamma_R}\Big(  \text{Im} G^R(\omega) -\overline{ \dot x^2} \; \text{Im} \Big[ \Phi   \Big( G^R(\omega) \Big)^4  + 3 \Lambda^2 \Big( G^R(\omega) \Big)^5 \Big]\Big).
\label{t2}
\end{equation}
Here the first term is the standard transmission coefficient for elastic electron transport computed for frozen molecular geometry\cite{haug-jauho}
and the second term accounts for the nonadiabatic corrections due to time-dependent variations  of molecular geometries.
The simple mathematical analysis of (\ref{t2}) tells us that the nonadiabatic effects are mostly significant in the resonance regime; when the system is out of resonance the higher powers of the retarded Green's function become small due to the presence of the $\omega - \epsilon$ term in the denominator. We also see that the $\Lambda^2$ term in (\ref{t2}) gives a major contribution to the non-adiabatic correction as the system approaches the resonance regime since  $\text{Im} \; \Big( G^R(\omega) \Big)^5 \gg \text{Im} \; \Big( G^R(\omega) \Big)^4$ if $\omega$ is in the vicinity of $\epsilon$. 

\begin{figure}[t!]
\begin{center}
\includegraphics[width=1.0\columnwidth]{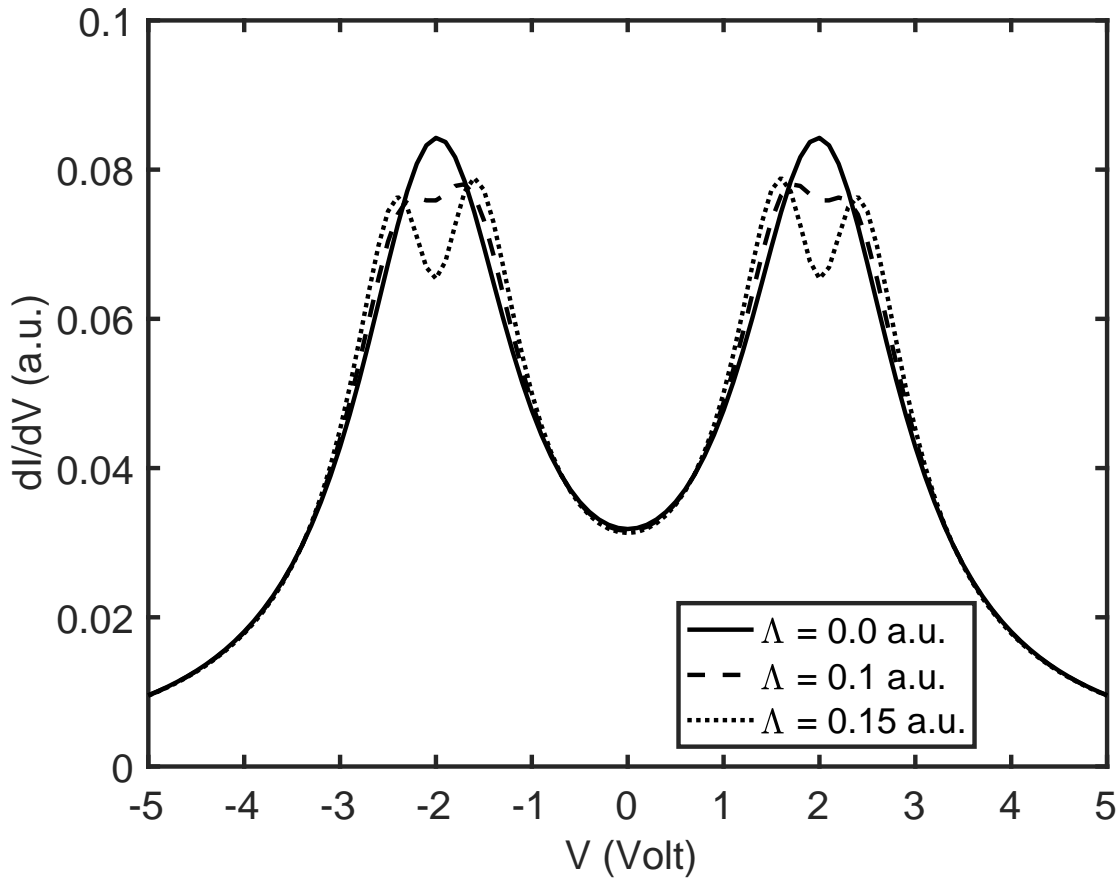}
\end{center}
	\caption{Conductance as a function of applied voltage computed for different values of $\Lambda$. Parameters used in calculations: $\Gamma_L=\Gamma_R=0.5$ eV, $\epsilon=-1$ eV, $\Phi=0.1$ a.u., $\overline{ \dot x^2 } = 4\times 10^{-7}$ a.u.}	
\label{figure1}
\end{figure}
Exactly at the resonance $\omega=\epsilon$, (\ref{t2}) becomes
\begin{equation}
{\cal T}(\epsilon) =  \frac{2}{\pi}  \frac{\Gamma_L \Gamma_R}{(\Gamma_L + \Gamma_R)^2}\Big(  1- \frac{ 48 \overline{ \dot x^2} \Lambda^2}{ (\Gamma_L + \Gamma_R)^4}\Big).
\label{t3}
\end{equation}

Therefore, the contribution from the $\Phi$ term completely disappears and nuclear motion plays a destructive role in resonance electron transport  always reducing the molecular conductivity.Physically ${\cal T}(\omega)$ must always be a positive quantity, therefore, the appearance of the negative value indicates that the more terms should be included into the power series expansion (\ref{expansion})  for the solution of the  equation of motion for the Green's function.

Using (\ref{t2}) we  calculate the adiabatic conductance and its non-adiabatic corrections. 
Let us, first, estimate physically meaningful values of molecular orbital derivatives $\Lambda$, $\Phi$  and average velocity $\overline{ \dot x^2 }$
which can be used in our calculations. We take a set of several representative diatomic molecules and compute $\Lambda$ and  $\Phi$ as derivatives with respect to the bond length of the highest occupied molecular orbital using Hartree-Fock theory with a 6-311+G(3df,3pd) basis set.\cite{gaussian09}
The result is: $\Lambda=-0.2$ a.u., $\Phi=0.3$ a.u. for O$_2$,   $\Lambda=0.2$ a.u, $\Phi=-0.11$ a.u for H$_2$ and $\Lambda=-0.1$ a.u., $\Phi=0.01$ a.u. for CO.  We also estimate typical values of the $\overline{ \dot x^2 }$ using experimental values of diatomic vibrational frequencies and relating average kinetic energy to half of the molecular zero-point energy. Average velocity  $\overline{ \dot x^2 }$  lays in the range of $10^{-7}$ a.u. for most diatomics.  Therefore, we see that nonadiabatic transmission coefficient  (\ref{t2},\ref{t3}) retains positive values for all  physically reasonable values of the molecular junction parameters.

We plot the conductance as a function of applied voltage as seen in Figure 1 for various typical values of $\Lambda$. We see that the nonadiabatic effects are mostly pronounced in  the resonant regime of electron transport. The resonance conductance peaks are suppressed and split by nuclear motion.

\section{Conclusion}
In this paper, we have developed quantum transport theory which takes into account  nonadiabatic effects of  nuclear motion. Our approach was based on Keldysh nonequilibrium Green's functions and the use of Wigner representation to solve the equations of motion for the Green's functions. The use of  Wigner representation enabled the separation of  fast and slow timescales. For the Green's function ${\cal G}$, the slow nuclear motion implied that ${\cal G}(T+\tau/2,T-\tau/2)$ varies slowly with the central time $T$, but oscillates fast with the relative time $\tau$. Wigner representation integrated out the fast time  $\tau$ to the energy domain and treats the slow-varying time $T$
explicitly.  We used the time derivatives with respect to central time  in the  Green's function as a small parameter and developed systematic perturbative expansion to solve the equations of motion for the Green's functions in Wigner space.  We produced analytic expressions  for nonadiabatic electronic Green's functions
  that depend on the instantaneous molecular geometry,  nuclear velocities and accelerations.  
 The general expression for the electric current in terms of Green's functions and self-energies was converted to Wigner space maintaining terms up to the second order in the central time derivatives. Combining Wigner space equation for electric current with the nonadiabatic nonequilibrium Green's functions we obtained nonadiabatic corrections to standard formula for the electric current.  Our method was not based on the assumption that the amplitude of nuclear motion is small or harmonic, nor it assumed that the coupling between nuclear and electronic degrees of freedom is small.  
 The extended formula for electric current   was cast into a form suitable for the implementation into existing density functional theory codes for electron transport calculations; the corrections to the electric current was expressed in terms of adiabatic Green's functions and nuclear gradients of molecular orbital energies and all of these ingredients being readily available. 
  
The details of the theory were further explained by application to the electron transport in a molecular junction with single resonant molecular orbital where calculations were performed in the wide-band approximation. In this case, the expressions for nonadiabatic Green's functions and electric current become transparent revealing the physical meaning of each term.   It was shown analytically that nonadiabatic effects play destructive role in molecular conductivity for electron transport in the resonance regime due to growth of higher order retarded Green's function in the nonadiabatic term.

\newpage
\appendix
\section{Equations of motion in Wigner representation}
The time derivative is expressed using central  $T$ and difference  $\tau$  times   as
\begin{equation}
\partial_{t} = \frac{1}{2}\partial_T + \partial_{\tau}.
\end{equation}
Conversion to the Wigner representation is done by considering the LHS and RHS of  equation of motion  (\ref{eom1}) separately. 
First, we consider the LHS 
\begin{equation}
A(t,t') = \left[ i\partial_t - h \left( \textbf{R}(t) \right) \right] {\cal G }^R(t,t'),
\end{equation}
which in terms of central and difference time becomes
\begin{equation}
A \left( T + \frac{\tau}{2},T - \frac{\tau}{2} \right) = \left[ \frac{i}{2}\partial_T + i\partial_{\tau} - h \left( \textbf{R}(T + \frac{\tau}{2}) \right) \right] {\cal G }^R \left( T + \frac{\tau}{2},T - \frac{\tau}{2} \right),
\label{a}
\end{equation}
By applying the Wigner transform to (\ref{a}) and expanding the brackets we find that 
\begin{eqnarray}
\tilde{A}(T, \omega) =  \int^\infty_{-\infty} d\tau \; e^{i\omega\tau} \Big(\frac{i}{2} \partial_T + i \partial_\tau \Big) {\cal G }^R \left( T + \frac{\tau}{2},T - \frac{\tau}{2} \right)   
\nonumber
\\
- \int^\infty_{-\infty} d\tau \; e^{i\omega\tau}h \left( \textbf{R}(T + \frac{\tau}{2}) \right) {\cal G }^R \left( T + \frac{\tau}{2},T - \frac{\tau}{2} \right). 
\label{aW}
\end{eqnarray}
The first term in (\ref{aW})  can be easily computed using integration by parts and taking into account that ${\cal G }^R \rightarrow 0$  if $|\tau| \rightarrow \infty$. It simply becomes
$(\frac{i}{2}\partial_T +\omega) \widetilde{{\cal G }}^R(T,\omega)$.  The calculation of second integral is discussed in some details below:
\begin{align}
& \int^\infty_{-\infty}  d\tau e^{i\omega\tau}h \left( \textbf{R}(T + \frac{\tau}{2}) \right) {\cal G }^R \left( T + \frac{\tau}{2},T - \frac{\tau}{2} \right) 
= \int^\infty_{-\infty}  d\tau  e^{i\omega\tau} e^{\frac{\tau}{2} \partial_T^h}h \left( \textbf{R}(T) \right) {\cal G }^R \left( T + \frac{\tau}{2},T - \frac{\tau}{2} \right)
\nonumber
\\
& 
= \int^\infty_{-\infty} d\tau e^{\frac{1}{2i} \partial_\omega \partial_T^h}   e^{i\omega\tau}  h \left( \textbf{R}(T) \right) {\cal G }^R \left( T + \frac{\tau}{2},T - \frac{\tau}{2} \right)  
= e^{\frac{1}{2i} \partial_\omega \partial_T^h} h \left( \mathbf R(T) \right)\; \widetilde{{\cal G }}^R(T,\omega).
\end{align}
Combining we find that $\tilde{A}(T, \omega)$ is given by 
\begin{equation}
\tilde{A}(T, \omega) = \left[ \omega + \frac{i}{2}\partial_T  - e^{\frac{1}{2i}\partial_{\omega} \partial_{T}^{h}} h \left( \textbf{R}(T) \right) \right] 
\tilde{{\cal G }}^R(T,\omega).
\end{equation}

We now consider the RHS of the Dyson equations
\begin{equation}
B(t, t') = \delta(t-t') +  \int_{-\infty}^{+\infty} dt_1 \Sigma^{R}(t,t_{1}){\cal G }^{R}(t_{1},t'). 
\label{b}
\end{equation}
Applying the Wigner transform to (\ref{b}) and taking note of the following identity  for the Wigner transformation of the convolution of two functions\cite{haug-jauho}
\begin{equation}
\int_{-\infty}^{+\infty}   e^{i\omega\tau}C(t,t_{1})D(t_{1},t') d\tau = e^{\frac{1}{2i} \left[ \partial_{T}^C \partial_{\omega}^D- \partial_{\omega}^C \partial_{T}^D \right] }\widetilde{C}(T,\omega)\widetilde{D}(T,\omega),
\label{a9}
\end{equation} 
which when implemented to (\ref{b}) with the knowledge that the self energy does not depend  on central time gives us
\begin{equation}
\widetilde{B}(T, \omega) = I + e^{-\frac{1}{2i}\partial_{\omega}^{\Sigma} \partial_{T}^{{\cal G }}} \; \Sigma^{R}(\omega)\widetilde{{\cal G }}^{R}(T,\omega). 
\end{equation}
Thus by incorporating the computed LHS and RHS  we find that the equation of motion in Wigner representation is given by (\ref{eomW}).

\section{Averaged quantities for electric current}

The second order correction to the retarded Green's function (in Wigner representation) averaged over nuclear velocities is
\begin{eqnarray}
\nonumber
 &&\overline{ \tilde{G}_{(2)}^{R} }=   \overline{ \dot{x}_\alpha \dot{x}_{\beta}}  \Big\{ - \frac{1}{8} G^R   \Phi_{\alpha \beta}   G^R \Big(  2 {\cal A}  G^R  {\cal A}
 + \partial_{\omega}^{2} \Sigma^R \Big) G^R
+ \frac{1}{4}  G^R \Big[ {\cal A} G^R , \Lambda_\beta  G^R \Big] \Big[ \Lambda_\alpha  G^R, {\cal A} G^R \Big]  
\\
\nonumber
 &&+ \frac{1}{4}  G^R {\cal A} G^R \Big[ \Big( \Phi_{\alpha \beta} + \Lambda_{\alpha}  G^R \Lambda_\beta \Big)  G^R, {\cal A}  G^R \Big] 
+  \frac{1}{4}  G^R {\cal A}  G^R \Big[ \Lambda_{\alpha} G^R, {\cal A} G^R \Lambda_\beta G^R \Big] 
\\
&&
 - \frac{1}{4}  G^R \Lambda_{\alpha}  G^R \Big[ \Lambda_{\beta} G^R  {\cal A} G^R, {\cal A}  G^R \Big]  
 - \frac{1}{4}  G^R \Lambda_{\alpha}  G^R \Big[ \Lambda_{\beta} G^R, \big( \partial_{\omega}^{2}  \Sigma^R +  {\cal A} G^R  {\cal A} \big)  G^R \Big]
 \Big\}.
\end{eqnarray}
Time and frequency derivatives of the first order correction to the retarded Green's function (in Wigner representation) averaged over nuclear velocities is
\begin{multline}
 \overline{ \partial_{T} \partial_{\omega}    \widetilde {\cal G}^R_{(1)}  } = \frac{i}{2} \overline{ \dot{x}_\alpha \dot{x}_\beta } 
\hat R_\beta \Big\{
G^R {\cal A} G^R \left[ {\cal A }G^R, \Lambda_\alpha G^R \right] 
+ G^R\left[ {\cal A} G^R,  \Lambda_\alpha G^R {\cal A} G^R\right]
\\
+ G^R \left[ {\cal A} G^R { \cal A}  G^R - \partial_\omega^2 \Sigma^R G^R, \Lambda_\alpha G^R \right]
\Big\}
+ \frac{i}{2} \overline{ \dot{x}_\alpha \dot{x}_\beta } 
\Big\{
G^R {\cal A} G^R \left[ {\cal A} G^R, \Phi_{\alpha \beta} G^R \right] 
\\
+ G^R\left[ {\cal A} G^R,  \Phi_{\alpha \beta} G^R {\cal A} G^R\right]
- G^R \left[ \partial_\omega^2 \Sigma^R G^R, \Phi_{\alpha \beta} G^R \right]
\Big\}.
\end{multline}
Here and below  $\hat R_\beta$ is the "replacement" operator which replaces matrix $G^R$ one-by-one by product of three matrixes ${G}^{R} \Lambda_\beta {G}^{R}$. For example,
$\hat R_\beta \{ G^R {\cal A } G^R \}$ means $ G^{R} \Lambda_\beta {G}^{R}{\cal A} G^R + G^R {\cal A} {G}^{R} \Lambda_\beta {G}^{R}$.
Time and frequency second derivatives of the adiabatic retarded Green's function  averaged over nuclear velocities
\begin{multline}
\overline{ \partial^2_{T} \partial^2_{\omega}    G^R  } =  \overline{ \dot{x}_\alpha \dot{x}_\beta } 
\hat R_\beta 
\Big\{
G^R \Lambda_\alpha  G^R \Big( 2 {\cal A} G^R {\cal A} + \partial^2_\omega \Sigma^R \Big) G + G^R  \Big( 2 {\cal A} G^R {\cal A} + \partial^2_\omega \Sigma^R \Big) G^R \Lambda_\alpha  G^R 
\\ 
+2 G^R {\cal A} G^R {\cal A}  G^R \Lambda_\alpha  G^R 
\Big\}
+
 \overline{ \dot{x}_\alpha \dot{x}_\beta } 
\Big\{
G^R \Phi_{\alpha \beta} G^R \Big( 2 {\cal A} G^R {\cal A} 
+ \partial^2_\omega \Sigma^R \Big) G 
\\
+ G^R  \Big( 2 {\cal A} G^R {\cal A} + \partial^2_\omega \Sigma^R \Big) G^R \Phi_{\alpha \beta}  G^R  +2 G^R {\cal A} G^R {\cal A}  G^R \Phi_{\alpha \beta}  G^R 
\Big\}.
\end{multline}

\clearpage
%

\end{document}